\documentclass[usegraphicx,usedcolumn,useAMS,usenatbib]{mn2e}

\title[Dynamical Origin of the ICM Metallicity Evolution]{On the Dynamical Origin of the ICM Metallicity Evolution}

\author[S. A. Cora et al.]{S. A. Cora$^{1,2}$\thanks{E-mail:sacora@fcaglp.unlp.edu.ar},
L. Tornatore$^{3,6}$,
P. Tozzi$^{4,5,6}$ and K. Dolag$^{7}$\\
$^{1}$Facultad de Ciencias Astron\'omicas y Geof\'{\i}sicas de la Universidad 
Nacional de La Plata and Instituto de Astrof\'{\i}sica de La Plata,\\ 
Observatorio Astron\'omico, 
Paseo del Bosque S/N, 
1900 La Plata, Argentina\\
$^{2}$Consejo Nacional de Investigaciones Cient\'{\i}ficas y T\'ecnicas,
Rivadavia 1917, Buenos Aires, Argentina\\
$^{3}$ SISSA - International School for Advanced Studies, via Beirut 4, I-43100 Trieste, Italy\\
$^{4}$ Dipartamento di Astronomia dell'Universita di Trieste, via G.B. Tiepolo 11, I-34131, Trieste, Italy\\
$^{5}$ INFN - Istituto Nazionale di Fisica Nucleare, Trieste, Italy\\
$^{6}$ INAF - Istituto Nazionale di Astrofisica, Trieste, Italy\\
$^{7}$ Max-Planck-Institut f\"ur Astrophysik, Karl-Schwarzschild-Strasse 1,
85740 Garching bei M\"unchen, Germany
} 
\begin{document}

\date{}

\pagerange{\pageref{firstpage}--\pageref{lastpage}} \pubyear{2007}

\maketitle

\label{firstpage}

\begin{abstract}
We present a study on the origin of the metallicity evolution
of the intra-cluster medium (ICM)   
by applying a semi-analytic model of galaxy formation to
{\em N}-body/SPH (smoothed particle hydrodynamic)
non-radiative numerical simulations of clusters of galaxies.
The semi-analytic model includes gas cooling, star formation, supernovae 
feedback and metal enrichment, and is linked to the 
diffuse gas of the underlying simulations so that the chemical properties
of gas particles are dynamically and consistently generated from stars in the 
galaxies. This hybrid model let us have information on 
the spatial distribution of metals in the ICM.
The results obtained for a set of clusters with virial masses
of $\sim 1.5 \times 10^{15}\,h^{-1}\,{\rm M}_{\odot}$
contribute to the theoretical interpretation of recent observational {\em X}-ray
data, which indicate a decrease of the average iron content 
of the intra-cluster gas with increasing redshift.
We find that this evolution arises mainly as a result of
a progressive increase
of the iron abundance within $\sim 0.15\,R_{\rm vir}$.
The clusters have been considerably enriched by $z\sim 1$ with
very low contribution from 
recent star formation.
Low entropy gas that has been
enriched at high redshift
sinks to the cluster centre contributing to the
evolution of the metallicity profiles.

\end{abstract}

\begin{keywords}
galaxies: clusters: general - methods: numerical - cosmology: observations - {\em X}-rays: galaxies: clusters
\end{keywords}

\section{Introduction}\label{sec_intro}

The intra-cluster medium (ICM) is a hot ($\sim 10^7$~K) and diffuse gas 
contained within 
the deep potential well of clusters of galaxies, which constitute
the largest virialized structures of the Universe.
This gas radiates energy through thermal Bremsstrahlung, which is detected
in the X-ray band of the electromagnetic spectrum,
characterized by the presence of emission lines from highly ionized iron
(Fe XXV and Fe XXVI) at 6.6 - 7 keV. These lines allow to obtain information
about the metallicity of the ICM; 
the average Fe
abundance is measured to be around 0.5$Z_{\odot}$ 
in the central regions of
local X-ray clusters.\footnote{
We note that this value is different from the one found in the
literature, which is based on the photosferic solar Fe abundance
published in \citet{anders89}. We are using the more recent
solar value of iron abundance by number 
(${\rm Fe}/{\rm H})_{\odot}=2.82\times10^{-5}$ by \citet{asplund05}.}

Supernovae explosions are the typical sources of metals
that contaminate the ICM, being generated 
through subsequent episodes of star formation 
in cluster galaxies (e.g., \citealt{renzini97}),
although other sources may have a non negligible contribution
to the metal budget of the ICM, like
hypernovae associated with population type III
stars \citep{loewenstein01} and intra-cluster stars \citep{zaritsky04}.
The metals produced enrich the ICM by different 
processes that affect the heavy element distribution, like 
galactic winds (\citealt{heckman00}, \citealt{springel03}), 
ram pressure stripping \citep{kapferer07}, and active galactic
nuclei \citep{moll07}. 
The knowledge of the history of the chemical enrichment of the ICM
contributes to our understanding on the way and epoch in which the
galaxy clusters have formed, the history of star formation in cluster
galaxies and the associated supernovae rates, and the physical
processes involved in the diffusion of energy and metals within the
hot intra-cluster gas.
Observations of radial abundance profiles of different elements
(\citealt{degrandi01}; \citealt{tamura04}; \citealt{vikhlinin05})
provide valuable constraints on the physical
processes involved in the chemical enrichment of the ICM, 
especially those related to the 
feedback
mechanisms that inject metals into the diffuse phase.

A recent analysis of {\em Chandra} X-ray spectra of 56 clusters within
the redshift range $0.3 \la z \la 1.3$ spanning temperatures $3 \la
k\, T \la 18$~keV allows to trace the evolution of the iron content
of the ICM \citep{balestra07}.  The results are based on the
estimated average emission-weighted (EW) iron abundance of the ICM
within the inner region of the cluster, delimited by $\sim 15 - 30$ per cent of
the virial radius, $R_{\rm vir}$.  They
find that, for $z \ga$ 0.5, the mean abundance of the ICM is
approximately constant, with a value of $Z_{\rm Fe}\approx 0.4 \, Z_{\odot}$.
However, at lower redshifts ($0.3 \la z \la 0.5$), the EW 
iron abundance is significantly higher, reaching a value of
$Z_{\rm Fe} \approx 0.64\,Z_{\odot}$.  The parametrization of these results
implies that the iron abundance at the present epoch is a factor of
$\sim 2$ larger than at $z \simeq$ 1.2. 
These results are supported by the 
more recent study made by \citet{maughan08}
based on a larger sample of galaxy clusters at $0.1 \la  z \la 1.3$ observed with
{\em Chandra} ACIS-I. 

From the current observed ICM metallicity evolution, it is not
clear whether it arises as a result
of a change in the iron content of the intracluster gas
due to enhanced star formation, or as a consequence
of redistribution of metals within the central regions;
the later would be favoured by 
dynamical processes which transfer the chemically enriched gas
from the intergalactic medium of the cluster galaxies to the hot intracluster 
gas.

Since measurements of ICM abundances are emission-weighted,
the mean abundance of the intracluster gas contained within a given radius
is biased to higher values when considering the central regions
of the clusters.
This bias is expected to be higher in cool core clusters because of the
peaks in their surface brightness and iron abundance that characterize them
(e.g., \citealt{degrandi01}).
However, the
observed evolution of the ICM iron abundance cannot be entirely attributed to
a possible decrease of the fraction of cool-core clusters at high redshift
(\citealt{balestra07}, \citealt{maughan08}).

Some studies on the origin of the central peak in the metal distribution of
clusters with a cool core suggest that it has been likely produced by the
brightest cluster
galaxy after the cluster was assembled
(\citealt{bohringer04}, \citealt{degrandi04}). For this mechanism to be
effective, large enrichment times ($4 - 10$~Gyrs) are necessary.
This conclusion has been achieved from the combination of observational
results and population synthesis and chemical enrichment models.
This possible mechanism for the formation of the central peak
has been complemented with the effect
of the AGN-induced flows as a possible transportation
processes that have mixed the metals into the ICM
(\citealt{rebusco05}; \citealt{roediger07}), trying to explain
the observed difference between the broad ICM metallicity profiles and the
much narrower stellar light
profile of the central galaxy.

Different theoretical approaches have been used so far
aimed at studying the processes of metal
enrichment of the ICM.
By using observed and modelled rates of type Ia supernovae
(SNe Ia) and core-collapsed supernovae (SNe CC), 
\citet{ettori05} has evaluated their contribution
to the history of metal accumulation in the ICM, finding 
that they provide 
a good qualitative agreement with observations
of the overall decline in abundance with redshift
thanks to the large delay times for SNe Ia adopted.
However, the
total amount of iron is systematically lower than observed values 
in the redshift range $0 < z <1$. 
This model indicates that
half of the iron mass observed locally is produced by SNe Ia,
with SNe CC products becoming dominant at higher redshifts.
Different conclusions are obtained by  \citet{loewenstein06}
who uses a similar kind of model, implying instead that
synthesis of cluster iron was dominated by SNe CC and/or
SNIa with short delay times whose progenitors originated during a 
phase of rapid, top-heavy star formation.  

The impact on the ICM chemical evolution
of environmental effects, such as ram pressure, and tidal or 
viscous stripping has also been analyzed. 
Calura, Matteucci \& Tozzi (2007) use chemical evolution models 
applied to the different
morphological
types of cluster galaxies. They find that iron enriched gas
ejected by galactic winds arising from ellipticals accounts
for the ICM iron abundances at $z\ga0.5$, whereas the gas stripped
from the disk galaxies turning into S0 through 
interaction of the interstellar mediun with the ICM
accounts for the ICM metallicity at $z\la0.5$. 
The joint contribution of these processes
can explain 
the increase in the ICM abundance in the low redshift range observed by
\citet{balestra07}. 
By using a complete different approach based on
a combination of {\em N}-body 
with a phenomenological galaxy formation model,
which provides the ejected galactic matter that
is then included into a hydrodynamic ICM simulation, 
\citet{kapferer07}
find that the mass loss triggered by ram pressure stripping
is more dominant than galactic winds since a higher redshift ($z\la2$);
however, their enrichment model cannot account for the
observed evolution of the the mean ICM metallicity with redshift.

From an hybrid model that combines an {\em N}-body/SPH
non-radiative simulation of a galaxy cluster with a semi-analytic model
of galaxy formation, \cite{cora06} has shown that
dynamical effects play a major role in the development of
central iron abundance peaks. In the proposed scenario,
the intergalactic medium of infalling cluster galaxies, that has been
primarily enriched at
high redshifts and then transfered to the ICM, is
subsequently driven to the cluster centre by bulk motions in the
intracluster gas.
                                                                                
Using cosmological hydrodynamical simulations, \citet{tornatore07}
also find that
gas dynamical effects, related to gas mixing, galactic winds and ram
pressure stripping, play an important role in determining
the distribution and relative abundance of different chemical species,
being the typical age of the ICM enrichment of $z\simeq 0.5$.
However, the influence of an excess of star formation at low redshift
is more important than the one obtained with the hybrid model of
\citet{cora06}, which gives quite low recent star formation
for cluster galaxies.

For the theoretical interpretation of the observational results 
concerning the ICM metallicity evolution, we apply here
the hybrid model of the chemical enrichment of the ICM implemented by
Cora (2006).  
Although this hybrid approach does not couple metal
production to the gas actually cooling in radiative simulations (e.g.,
\citealt{tornatore07}), its novel feature is that the chemical
properties of the diffuse gas in the underlying non-radiative
{\em N}-body/SPH
cluster simulations are
generated from metals ejected from the
galaxies, consistent with the modelling of the semi-analytic
model. These metals are then carried around and mixed by the
hydrodynamic processes during cluster formation,
allowing us to study the evolution of the spatial distribution of metals in
the ICM.
We compare the results of our hybrid model with the mean iron abundances of
the ICM recently determined by \citet{balestra07}, with the aim of
investigating the connection between the development of the chemical
abundance patterns that characterize the ICM and the observed average
metallicity evolution.

This paper is organised as follows.  Section~\ref{sec_model}
briefly describes the hybrid model used to study the ICM chemical enrichment,
presenting  the changes applied to the version described in 
\citet{cora06}, mainly related to the distribution of metals
among gas particles; it also 
summarises the properties of the hydrodynamical simulations used.
Section~\ref{sec_evol} contains the analysis of the evolution of the
ICM chemical enrichment considering both radial abundance profiles 
(Section~\ref{sec_evol_prof}) and mean metallicities 
(Section~\ref{sec_evol_mean}) of the main cluster progenitor at different 
redshifts.
In Section~\ref{sec_conclu} we summarize our conclusions.

\section[]{Model of the ICM chemical enrichment}\label{sec_model}

We consider an hybrid model for studying the chemical enrichment of the
ICM, which consists of a combination of  
non-radiative cosmological {\em N}-body/SPH simulations that contain
galaxy clusters and a semi-analytic model
of galaxy formation.
The important feature of this hybrid model 
is the link between semi-analytic model results and the chemical
enrichment of the diffuse gas component of the underlying {\em N}-body/SPH
simulation. 
We pollute gas particles with metals ejected from the
galaxies that have been generated by  
the semi-analytic
model. 
By enriching the gas particles locally around galaxies, we can account
for the spreading and mixing of metals by hydrodynamical processes,
thereby obtaining a model for the evolution of the spatial
distribution of metals in the ICM.

\subsection[]{Semi-analytic model of galaxy formation}\label{sec_SAM}

The semi-analytic model used here is based on previous works
(\citealt{springel01}; \citealt{lucia04}), but was
extended with a new chemical implementation that tracks the abundance
of different species produced by different sources.
We briefly describe here the modelling of the
circulation of metals among the
different baryonic components.
Details of this model are given in \citet{cora06}.
The simplicity of the semi-analytic model has the advantage of reaching a
larger dynamic range than fully self-consistent hydro-simulations, at
a far smaller computational cost. In particular, it allows us to
explore more easily the range of parameters that characterize
appropriate chemical enrichment models.
The long cooling time of the bulk of the gas in rich clusters
justifies the assumption of a non-radiative gas in the
{\em N}-body/SPH simulations.
However, this time-scale becomes smaller than the
Hubble time in the core of the cluster, where the densities are
higher.  Here is where the semi-analytic model plays an important
part, taking into account radiative cooling, star
formation, and chemical enrichment and energetic feedback from galaxies
by including the effect of supernovae explosions.

\begin{table}
  \caption{
Number of high resolution dark matter particles, $N_{\rm hr}$,
of the non-radiative cosmological hydro-simulations, and
the virial mass, $M_{\rm vir}$, temperature, $T_{\rm vir}$ and
radius, $R_{\rm vir}$, of the most massive cluster in each simulation.
The simulation name used in \citet{dolag05} is given between brackets.
}
  \begin{tabular}{@{}lcccc@{}}
  \hline
  Cluster  & $N_{\rm hr}$ &
$M_{\rm vir}$ & $k\,T_{\rm vir}$ & $R_{\rm vir}$  \\
   & &
 [$h^{-1}\,{\rm M}_{\odot}$]&  [keV] &  [$h^{-1}$~Mpc] \\
 \hline
 C1 (g51)    & $2219034$ &  $1.3 \times 10^{15}$& 7.8 & 2.3 
\\
 C2 (g1)     & $4937886$ &  $1.5 \times 10^{15}$& 8.4 & 2.4 
\\
 C3 (g8)    & $5602561$ &  $1.7 \times 10^{15}$& 8.1 & 2.8 
\\
\hline
\end{tabular}
\end{table}

Dark matter haloes and substructures that emerge
in the simulation are tracked by the semi-analytic code and used to
generate the galaxy population \citep{springel01}.
In this subhalo scheme, the virial mass of a dark matter
substructure is defined simply as the sum of the mass of its particles,
and the virial radius is estimated by assuming that the subhalo has an
overdensity of 200 with respect to the critical density.
Since each particle in the {\em N}-body/SPH simulations
is split into dark matter and gas, the
identification of dark matter haloes is based on the dark matter particles
with their masses corrected for the cosmic baryon fraction.
This is done to get haloes defined as close as possible to pure dark matter
runs, on which 
semi-analytic models are usually calibrated, but still allows to
make
use of the hydrodynamic treatment for the distribution of metals
ejected by galaxies.

We assume that the hot gas always has a distribution that
parallels that of the dark matter halo. Its mass is initially
given by the baryon fraction of the virial mass of the dark matter halo,
and subsequently modified by gas cooling, star formation and feedback
processes.
Stars can contaminate the cold and hot gas because of
mass loss during their stellar evolution and metal ejection at the end
of their lives.  The hot gas has primordial abundances initially
($76$ per cent of hydrogen and $24$ per cent of helium), but becomes chemically
enriched as a result of the direct ejecta from stellar mass
loss, and the transfer of contaminated cold gas to the
hot phase that is reheated by supernovae explosions.  This chemical
enrichment has a strong influence on the amount of hot gas that can
cool, since we are using metal dependent cooling rates \citep{suth93}. 
This process
in turn influences the star formation activity which is ultimately
responsible for the chemical pollution.

The chemical model implemented in our semi-analytic code considers mass 
losses from stars in different
mass ranges.
Massive stars
give raise to core collapse supernovae,
which include those of type Ib/c and II, being the later 
the most abundant ones.
We adopt stellar yields as given from models of
\citet{marigo01} for low- and
intermediate-mass stars $(0.8 {\rm M}_{\odot} \la M \la 5 - 8\, {\rm M}_{\odot})$ and
\citet{portinari98} for quasi-massive 
$(5\, {\rm M}_{\odot} \la M \la 8\,
{\rm M}_{\odot})$ and massive stars $(8\, {\rm M}_{\odot} \la M \la
120\, {\rm M}_{\odot})$.
Ejecta from type Ia supernovae are also included, considering 
the nucleosynthesis
prescriptions from the updated model W7 by \citet{Iwamoto99}.
We use the stellar lifetime given by \citet{padovani93} to model
the return time-scale of the ejecta from all sources considered,
being specially relevant for the single stars in the
low- and intermediate-mass range
and for the progenitors of SNe Ia,
that are characterized by a long delay time from the formation of the progenitor
to the supernova explosion.
We adopt the single degenerate model to estimate the SNe Ia rate,
following the scheme of \citet{greggio83}, where type Ia supernovae originate
in binary systems whose components have masses between
$0.8$ and $8 \,{\rm M}_{\odot}$. Calculations are based on
the formalism described in \citet{lia02}, but assuming that
core collapse supernovae,
originate from single stars with masses
larger than $8\,{\rm M}_{\odot}$.

The mass range of secondary stars in binary systems
gives explosion times for SNe Ia comprised between
$\sim 2.9 \times 10^{7}$ and  $\sim 1.4 \times 10^{10}$~yrs, 
with the SNe Ia rate
reaching a maximun within $\sim 0.1 - 0.7$~Gyr for a single stellar population.
Thus, the ejection rate of
elements produced by SNe Ia reaches a maximum at $z \sim 3 - 5$, 
while the contribution of SNe CC to the ICM chemical enrichment peaks at higher redshifts ($z \sim 5 - 7$) since SNe CC rate closely follows
the star formation
rate as a result of the short lifetimes of the stars involved.
The peaks in the mass ejection rates
are followed by a strong decline
at lower redshifts for both types of SNe,
such that the ongoing chemical contamination is quite low 
at $z=0$ \citep{cora06}.
The relative contribution of SNe CC and SNe Ia to the iron
content of the cluster is already one-third
at $z\sim 2$, as given by
the accumulated iron masses provided by both sources.

Some aspects of the semi-analytic model have been modified with respect to
the version described in \citet{cora06}.
We consider here a Salpeter IMF normalized between 0.1 and
$100 \,{\rm  M}_{\odot}$. 
In order to obtain a mass-metallicity relation whose evolution
with redshift approaches to the observed one \citep{erb06},
half of the mass ejected by galaxies because of mass loss through 
stellar winds
or supernovae
explosions are transfered directly to the hot phase instead
of being first
incorporated to the cold gas.
In this way, the cold gas associated to each galaxy
is more gradually contaminated, 
instead of already achieving at high
redshifts the higher metallicitiy values that correspond to the
present epoch.

The free parameters of the model
regulate the way in which gas cooling, star formation, supernovae
feedback and galaxy mergers proceed, and
determine the circulation of metals
among the different baryonic components.
They
have been carefully tuned in \citet{cora06}
to satisfy numerous observational
constraints (Milky Way properties, the luminosity
function, the Tully-Fisher, color-magnitude and mass-metallicity relations).
The feedback efficiency parameter, that regulates
the amount of cold gas that is reheated by 
core collapse supernovae,
is now assigned a smaller value of 0.1 that allows to recover
a better agreement of model results with 
the observed slope of the Tully-Fisher relation.

\subsection[]{Non-radiative {\em N}-body/SPH cluster simulations}\label{sec_simu}

The semi-analytic model is applied to cosmological hydrodynamical
simulations.
We consider a set of three simulated galaxy clusters, having virial mass in
the range $\simeq (1-2)\times 10^{15}\,h^{-1}\,{\rm M}_\odot$.  These clusters
have been initially selected from a dark matter simulation of a cosmological
box, having size of $479\,h^{-1}$~Mpc \citep{yoshida01}, for a
cosmological model $\Omega_{\rm m}$=0.3, $\Omega_{\Lambda}$=0.7, ${\rm
H}_{\rm o}= 70 \, {\rm km} \, {\rm s}^{-1} \, {\rm Mpc}^{-1}$,
$\Omega_{\rm b}=0.039$ for the baryon density parameter, and
$\sigma_8=0.9$ for the normalization of the power spectrum.  The
Lagrangian regions surrounding the three selected clusters have been
re-simulated at higher mass resolution by applying the Zoomed Initial
Condition technique \citep{tormen97}. The mass resolution is the
same for all simulations, with masses of dark matter and gas particle
of $m_{\rm dm}=1.13 \times 10^{9}\,h^{-1}\,{\rm M}_{\odot}$ and $m_{\rm
gas}=1.69\times 10^{8}\,h^{-1}\,{\rm M}_{\odot}$, respectively. As for the
force resolution, the Plummer-equivalent gravitational softening is
fixed at $\varepsilon=5 \,h^{-1}$~Kpc in physical units at redshift
$z\le 5$, while it switches to comoving units at higher
redshifts. Table 1 gives the number of high resolution dark matter
particles of each simulation, which is equal to the number of gas
particles, and the virial properties of its most massive cluster.
The simulations have been carried out using the Tree-SPH {\tt
GADGET-2} code \citep{springel05}. {\small GADGET-2} is a parallel
Tree+SPH code with fully adaptive time-stepping, which includes an
integration scheme which explicitly conserves energy and entropy
\citep{springel02}. 
The simulations considered here include
only non-radiative physics and the original formulation of artificial viscosity
within SPH. General properties of these simulated clusters,
such as radial profiles of velocity dispersion of dark matter and
gas particles, and of gas temperature, density and entropy, are
given in \citet{dolag05}. 

\subsection[]{Hybrid model of the ICM metal enrichment}\label{sec_simu}

The key physical processes that lead to the evolution of the
iron content of the ICM are the production of metals and their deposition
in the gas. The former is controlled by the prescriptions included
in the semi-analythic model, while the later 
depends on the way in which 
chemical elements generated by
galaxies
in the semi-analytic model are distributed among gas particles
in the {\em N}-body/SPH cluster simulation, 
which thus carry the information on the spatial distribution of chemical and
thermodynamical properties of the ICM. 
The metal deposition results from a complex combination of different processes
that includes powerful outflows from star forming galaxies, 
generated by supernovae explosions
and winds from young massive stars,
dynamical stripping and gas mixing.  
We model the deposition of metals in a simple way.
For each snapshot of the simulation, we
identify a set of $N_{\rm gas}$ gas particles
within a sphere characterized by an enrichment radius, $R_{\rm enrich}$, 
centred on each galaxy.
Metals ejected by each galaxy are spread among the $N_{\rm gas}$ 
gas particles.
Both the
choice
of $R_{\rm enrich}$ and the
 procedure used to distribute the chemical elements
have been modified with respect to the version presented in 
\citet{cora06}.
In that work, metals were uniformly distributed within
a radius of $100 \, h^{-1}$~Kpc for all galaxies.

In the present implementation, the mass 
of iron ejected by the galaxy 
is distributed among its corresponding set of $N_{\rm gas}$ gas particles
according to the same kernel
used for SPH calculations in the {\em N}-body/SPH simulation,
that has been obtained from a {\em B}-spline.
Besides, we now consider
a physically motivated model for determining $R_{\rm enrich}$ 
based on a treatment of galactic winds 
(Bertone, Stoehr \& White 2005). 
It follows the expansion of supernova-driven 
superbubbles around several hundred thousand galaxies formed
in a region of space with diameter $52 \, h^{-1}$~Kpc and mean density 
close to the mean density of the 
universe. About half of the enclosed galaxies at $z=0$ are field galaxies, 
while the rest are in groups and poor clusters.
They solve the equation of motion for a spherical astrophysical blastwave
to follow the evolution of winds after they escape the visible regions
of galaxies.
Although the history of each wind depends on the properties of the parent 
galaxy and on the environment where the wind expands, \citet{bertone05}
find small scatter in the distribution of the shock radii for different
parameter choices of the model at a given redshift and as a function of the 
stellar mass of the galaxy. This 
ensures that mean quantities of global properties
of the winds well represent the general trend of the whole population.

Considering the results of \citet{bertone05}, we adopt the dependence of 
the mean bubble radius for all galaxies 
blowing a wind at 
$z=3$, as a function of the stellar mass, $M_{\star}$, of the host
galaxies (see their Figure 3). This dependence varies according to the
parameters chosen for the wind model, such that the bubble radius
ranges from $\approx 100$ to $400\, h^{-1}$~Kpc
for galaxies with $M_{\star}\approx 10^{11}\,{\rm M}_{\odot}$.
In order to explain the origin ot the observed evolution of the iron
content of the ICM, we first have to reproduce the 
iron spatial distribution at $z=0$.
We find that the curve for which 
galaxies 
with stellar mass $M_{\star}\approx 10^{11}\,{\rm M}_{\odot}$ have 
a bubble radius of $100\, h^{-1}$~Kpc
allows to recover
radial iron abundance profiles in good agreement with observations, 
as shown in Section~\ref{sec_evol_prof}. We fit this curve with  
a parametrization given by 
\begin{equation}
R_{\rm enrich}=-623.13+ 133.01 \, {\rm log} M_{\star} -6.13 \, {\rm log}^2 M_{\star}.\label{eqdist}
\end{equation}
Using a larger bubble radius for massive galaxies, or the same radius for 
all galaxies (fixed at $100\, h^{-1}$~Kpc as in \citealt{cora06}) give place
to radial iron abundance profiles flatter than observed.

It is important to note that we are not considering the evolution
of the bubble radius given by the treatement of \citet{bertone05}, 
but only the relationship between this radius with the stellar mass 
of the host
galaxy at a given redshift. The subsequent spreading and mixing of metals
is carried out by gas particles themselves through hydrodynamical processes.
The consistent implementation in our model of the equations
governing the wind
evolution and their impact on the chemical
pollution of the ICM will be presented in a forthcoming work.

\section[]{Evolution of ICM iron content}\label{sec_evol}

\subsection[]{Radial abundance profiles}\label{sec_evol_prof}

Observed iron abundances at different clustercentric radii
for local clusters (\citealt{tamura04,vikhlinin05}) constitute an important 
constraint to our model.
We evaluate its capability to represent the chemical
enrichment of the ICM by comparing
the radial iron abundance profiles of the ICM constructed from simulated
results at redshift $z=0$ 
with this kind of observational data.

Radial abundance profiles are built by dividing 
the volume limited by the cluster virial radius 
into concentric spherical shells 
centred on the dominant galaxy of the cluster. 
These shells have equal width till $0.1\,R_{\rm vir}$ and are 
logarithmically spaced beyond this radius.
For each shell, we estimate
the {\em X}-ray emission-weighted 
mean iron abundance relative to hydrogen,
using the emission measure of each gas particle defined as
$EM \equiv \int{n_{\rm p}\,n_{\rm e}\,dV}$,
where $n_{\rm p}$ and $n_{\rm e}$ are the proton and electron
densities, respectively, 
and $V$ is the volume associated to the
gas particle.
The iron abundance is
then expressed in terms of the solar value, for which we have adopted the
recent calibration of \citet{asplund05}.
These mean abundances are
assigned to the mean radius of each shell.

\begin{figure}
    \centering \includegraphics[width=80mm]{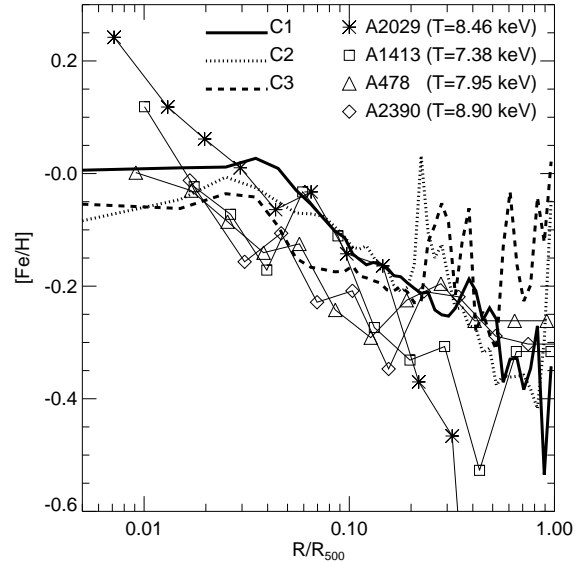}
\caption{
EW radial iron abundance profiles for the three simulated clusters
(thick solid, dotted and dashed lines).
Symbols represent 
observational data from \citet{vikhlinin05}.
}
\label{fig1}
\end{figure}

Figure~\ref{fig1} shows  
the EW radial iron abundance profile of the ICM at $z=0$ for 
the three simulated clusters (thick lines).
They are compared with a sample of nearby, relaxed galaxy
clusters observed with {\em Chandra} \citep{vikhlinin05},
with a range of average temperatures similar to that
of the simulated clusters.
The comparison of the model abundances with observed values is shown out
to $R_{500}$,
which is the radius enclosing a mean overdensity of $500$
with respect to the critical density at the redshift of the cluster.
The local EW profile shows a good agreement with the
trend traced  
by observations,
except in the innermost radial bins, being at odds
with self-consistent hydrodynamical simulations that include gas cooling.
The presence of dark matter substructures with associated high
density and chemically enriched gas
makes the
EW abundance profiles quite noisy at distances larger than
$\sim 0.2 \, R_{500}$
because of 
the strong dependence of the {\em X} - ray emissivity on the gas 
density. 
These peaks are not present in 
mass-weighted (MW) abundance profiles, which show almost the same trend than
EW ones but are much smoother.

\begin{figure}
    \centering \includegraphics[width=80mm]{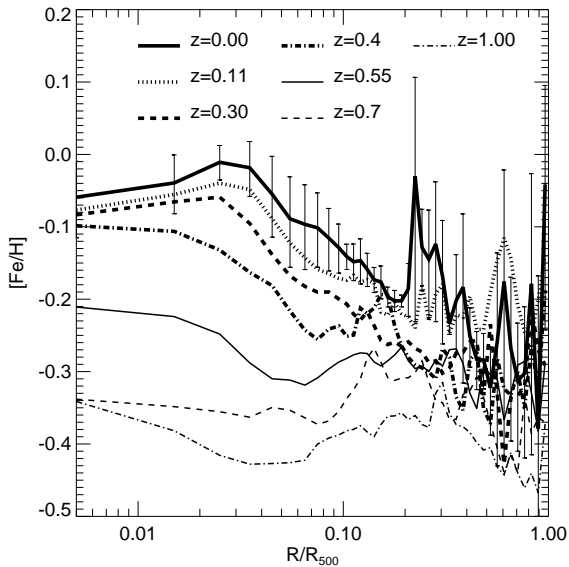}
\caption{
Evolution with redshift of EW radial iron abundance profiles.
They are
obtained by averaging
at each concentric shell the
abundance distribution of the three clusters analysed.
The error bars around the average radial profile at $z=0$ denote
the 1-$\sigma$ standard deviation of the mean. 
}
\label{fig2}
\end{figure}

The evolution of the radial abundance profiles is followed
by identifying the main progenitor of the cluster at different
redshifts. Then, we take into account the gas particles
that reside within the virial radius of these structures.
The main progenitor of a given halo at each redshift is selected
among the set of progenitors of that halo and defined as the
substructure
with the largest number of dark matter
particles.
Figure~\ref{fig2} shows the evolution with redshift
of EW radial iron abundance profile of the ICM.
Each curve corresponds
to a different redshift and has been obtained by
considering the gas particles 
within each radial bin for all the
three clusters, thus giving an average 
abundance distribution.
The thick solid line corresponds to
the present epoch, being the average of the individual radial
profiles represented  
by the thick curves in Figure~\ref{fig1}.

From Figure~\ref{fig2} we can appreciate that the EW 
iron abundances 
become progressively lower with increasing redshift,
with the main change occurring 
at radii less than 
$\sim 0.2\,R_{500}$ ($\sim 0.15\,R_{\rm vir}$), where the profiles become 
flatter.
At larger radii, the shape of the abundance distributions
at different redshifts keep themselves
quite close to
the profile at $z=0$. 
That is,
the evolution of the radial abundance profiles
from $z \sim 1$ to $z=0$ is mainly produced by 
an increment of the iron content in the innermost regions of the
clusters.

\subsection{Mean metallicity}\label{sec_evol_mean}

Recent results on the evolution of the iron content
of the ICM based on the analysis of {\em Chandra} {\em X}-ray spectra of 56
clusters of galaxies at $z \ga 0.3$ \citep{balestra07},
support a decrease of metallicity with redshift that can be parametrized
by a power law of the form $\sim (1+z)^{-1.25}$.
This observed evolution implies that the average iron content of the
ICM at the present epoch is a factor of $\sim 2$ larger than at
$z \simeq 1.2$.
Since this analysis is restricted to the
central regions ($R \simeq 0.15 - 0.3 \, R_{\rm vir}$) of clusters,
the evolution of the iron content might be
interpreted as a result of ocurrence
of evolution in the gradients of the metal
distribution. 
This hypothesis becomes possible taking into account our results on the
evolution of the EW radial iron abundance profiles shown in 
Figure~\ref{fig2}.

\begin{figure}
    \centering \includegraphics[width=80mm]{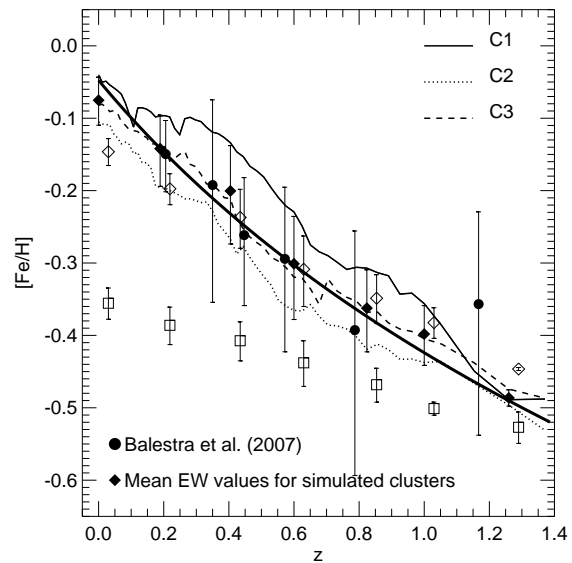}
\caption{
Evolution of the EW mean iron abundance of the main progenitors of the
 galaxy clusters analysed (thin lines).
These mean values have been estimated using
gas particles within $0.15 \, R_{\rm vir}$.
The mean values for all the three 
clusters are shown at some redshifts (filled diamonds; error bars denote
1$\sigma$ confidence level). The corresponding MW 
mean iron abundances
estimated within $0.15 \, R_{\rm vir}$ and $R_{\rm vir}$ are also shown 
(empty diamonds and empty squares, respectively).
These results are compared with the
observational data from \citet{balestra07} (filled circles)
and the corresponding parametrization of the metallicity evolution with
redshift (thick solid line). 
The observational data have been multiplied by a factor of
$1.66$ in order to refer the abundances to the solar values of
\citet{asplund05}.}
\label{fig3}
\end{figure}

In order to compare our results with those of
\citet{balestra07}, we estimate the mean EW
iron abundance
from the iron content of gas particles contained
within $0.15 \, R_{\rm vir}$
from the centre of the cluster or of its progenitor,
depending on the redshift considered.
Figure~\ref{fig3} shows this comparison for each of
the clusters analyzed, together with the corresponding mean
values for all the three clusters estimated at some redshifts
(filled diamonds). Observational data is represented by filled circles
and the thick solid line denotes the corresponding parametrization
of the metallicity evolution.
Model results closely follow the trend denoted by this parametrization
within the whole redshift range considered.
The innermost area contained within $0.15 \, R_{\rm vir}$ manifests the 
strongest changes 
in the slope of the radial abundance profiles (see Figure~\ref{fig2}).
Thus, the decrease of the observed EW mean iron abundances with 
increasing redshift
can be connected to the 
flattening of the central part of the abundance profiles. 

The rate of change of the EW mean metallicity
with redshift has a mild dependence with the region of the cluster considered,
manifesting clear evolution even for the case where 
all gas within $R_{\rm vir}$ is
taken into account (not shown in the figure). 
The low density and poorly contaminated gas particles lying at
distances greater than $\sim 0.15 \, R_{\rm vir}$ do not contribute much to the
EW mean metallicity because of their low {\em X}-ray emissivity.
Hence, EW mean abundances mainly
provide information of the central parts of the clusters.
 
Instead, estimations of MW mean iron abundances
give equal weight to both high and low enriched gas particles since
their masses do not change in a non-radiative SPH simulation.
Therefore, 
MW mean iron abundances strongly depend on the region of the cluster  
considered.
Taking into account gas particles within either 
$0.15 \, R_{\rm vir}$ or $R_{\rm vir}$ leads to MW mean metallicity
evolution 
with smaller values
when all gas contained within $R_{\rm vir}$ is involved in the calculations.
This is a consequence of the presence of progressively less
contaminated gas particles
with increasing clustercentric radius.
Results corresponding to $R_{\rm vir}$ and $0.15 \, R_{\rm vir}$ are shown
in Figure~\ref{fig3} 
by empty squares and empty diamonds, respectively.

As a result of the evolution of the slope of the radial abundance profiles
within $0.15 \, R_{\rm vir}$, 
the MW mean metallicity do show stronger evolution with
redshift only
when these inner parts of the cluster are considered, 
becoming quite close to the behaviour
given by EW values, as shown in 
Figure \ref{fig3}.
Hence, the observed metallicity evolution can be explained as being 
produced by the combination of two effects. On one hand, 
the evolution on MW mean iron abundances estimated from gas contained
within $R_{\rm vir}$ indicates that there is
an overall increase of the iron content
within the virial radius of the main progenitor of the cluster.
This is also evident from the lower values displayed by the whole
radial abundance profiles as redshift increases (see Figure~\ref{fig2}).
On top of this, we find that the evolution 
of the slope of the 
central part of the iron abundance profiles ($\la 0.15  \, R_{\rm vir}$)
is the key ingredient that contributes to achieve 
the observed evolutionary trend for $z \la 1.4$.

\begin{figure}
    \centering \includegraphics[width=80mm]{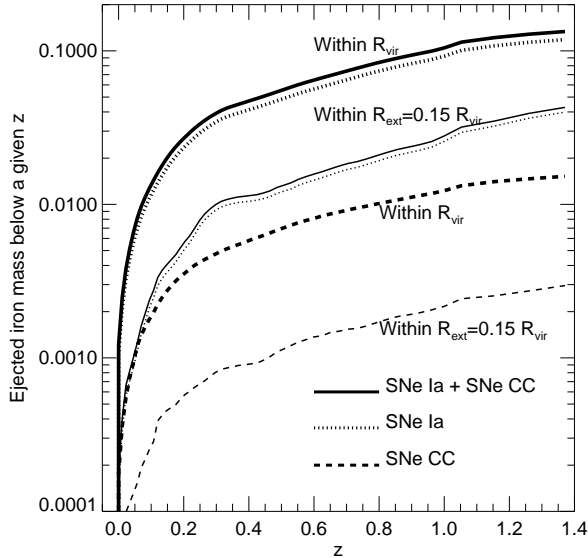}
\caption{
Evolution with redshift of
iron mass ejected by galaxies within $0.15\, R_{\rm vir}$ (thin lines)
and $R_{\rm vir}$ (thick lines) below a given redshift for cluster C1,
expressed in terms of the iron mass
contained within those radii at $z=0$.
Different types of lines indicate the joint or separate contribution of
supernovae type Ia and CC.}
\label{fig4}
\end{figure}

\subsection{Understanding the evolution of the ICM metallicity}\label{understand}

The joint evaluation of observational and simulated results
supports the fact that the observed evolution of the ICM iron content
till redshift $z \simeq 1.2$ is mainly due to the evolution of the
metallicity gradients of the intracluster gas
characterized by an increase of
central abundances with decreasing redshift.
Both the star formation rate and the mass accretion
history of the cluster may give raise to such evolution.
The study
carried out by \cite{cora06}, based on the analysis of a single simulated 
cluster, favours
the scenario in which
dynamical effects play the major role in the increment of the
central abundances of the ICM, since 
the amount of star formation at $z \la 0.5$ is not enough to account for
the observed evolution of the iron content.

\begin{figure*}
 \centering
  \begin{minipage}[c]{.33\textwidth}
    \centering \includegraphics[width=60mm]{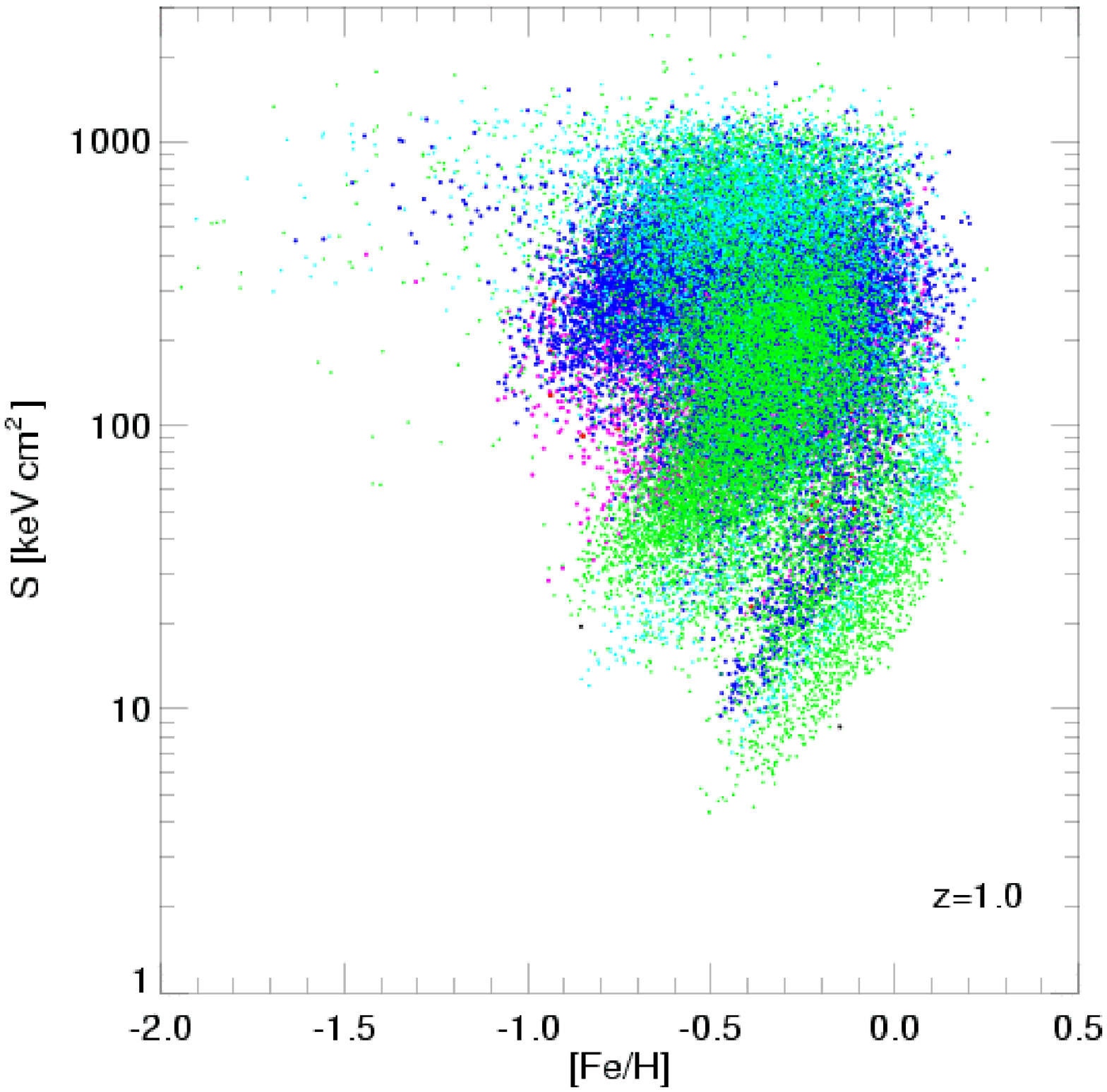}
  \end{minipage}%
  \begin{minipage}[c]{.33\textwidth}
    \centering \includegraphics[width=60mm]{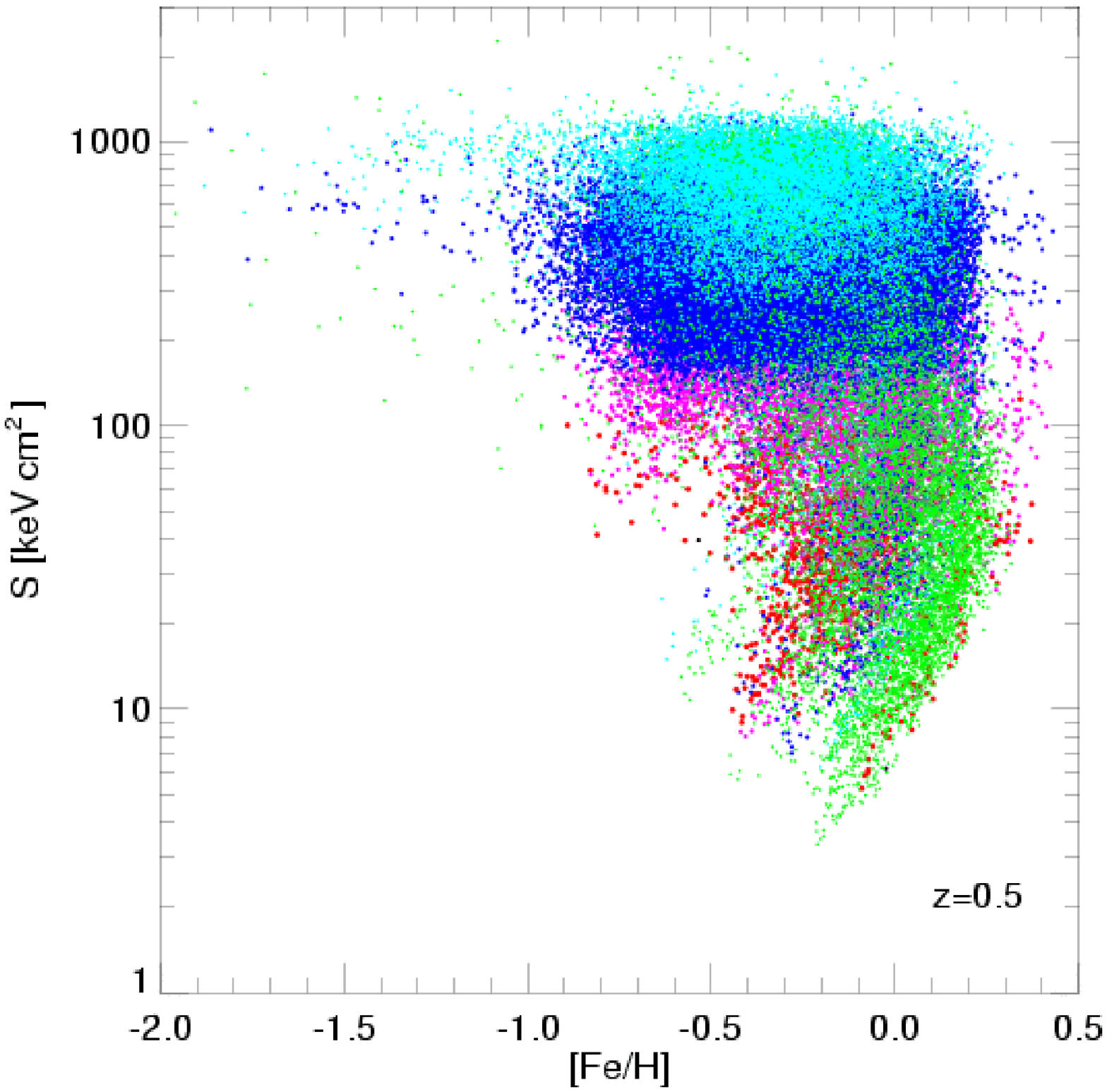}
  \end{minipage}%
  \begin{minipage}[c]{.33\textwidth}
    \centering \includegraphics[width=60mm]{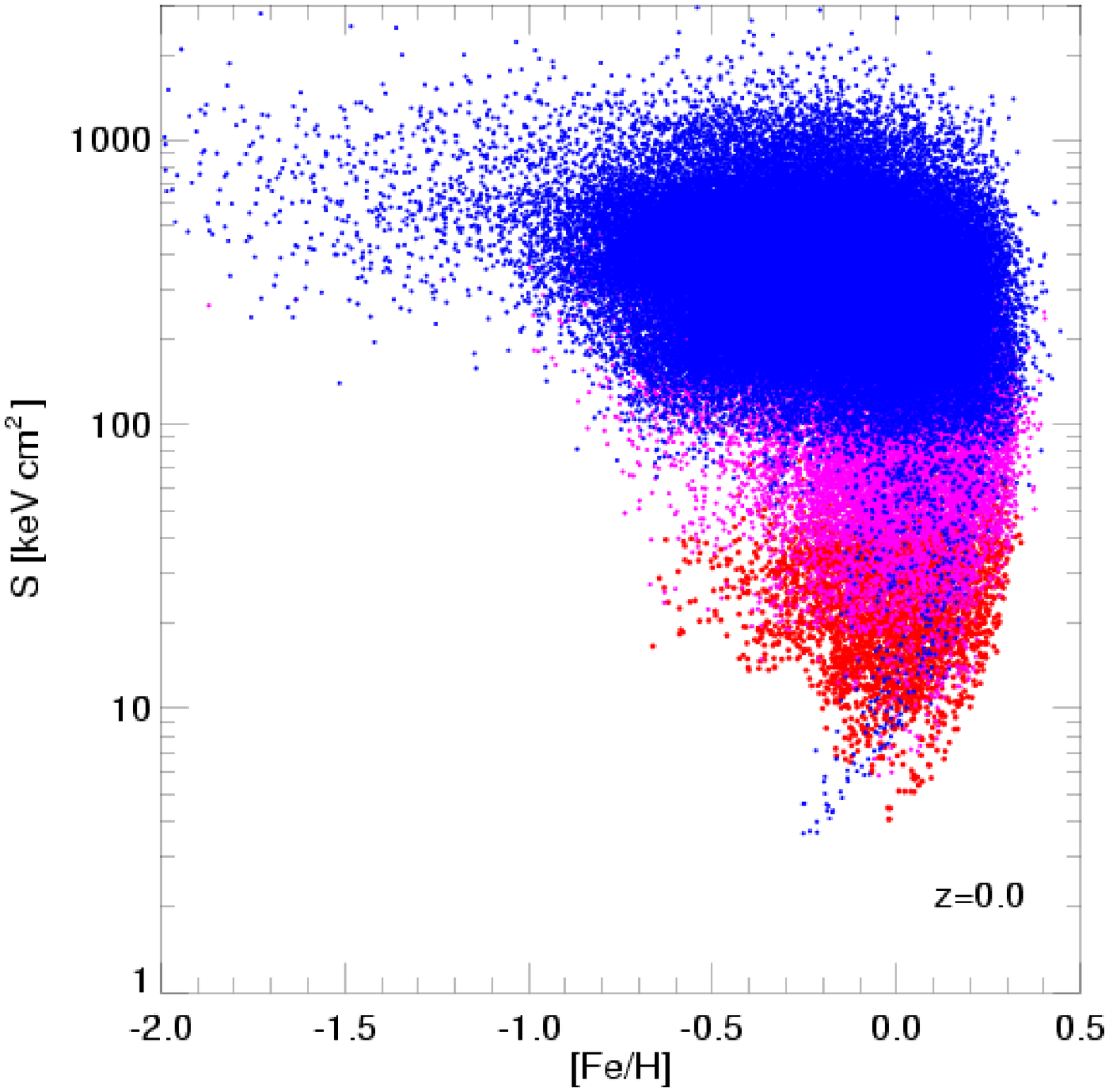}
  \end{minipage}%
\caption{Evolution with redshift of the dependence 
between the entropy and iron abundance
of  gas particles that have
distances from the cluster centre within a sphere of radius
 $0.15 \,R_{\rm vir}$ at $z=0$. Results correspond to 
cluster C1.
Redshifts are indicated in each panel.
Gas particles are colour-coded according to their distances 
with respect to the centre of the main progenitor of the cluster
at each redshift:
within  $0.02 \, R_{\rm vir}$ (red), in the ranges 
$0.02 - 0.05\, R_{\rm vir}$ (magenta),
$0.05 - 0.15\, R_{\rm vir}$ (blue),
$0.15 - 0.3 \, R_{\rm vir}$ (cyan) and $0.3 - 1 \, R_{\rm vir}$ 
(green). Gas particles outside $R_{\rm vir}$ are showed in Figure \ref{fig6}.  
}
\label{fig5}
\end{figure*}

\begin{figure*}
 \centering
  \begin{minipage}[c]{.33\textwidth}
    \centering \includegraphics[width=60mm]{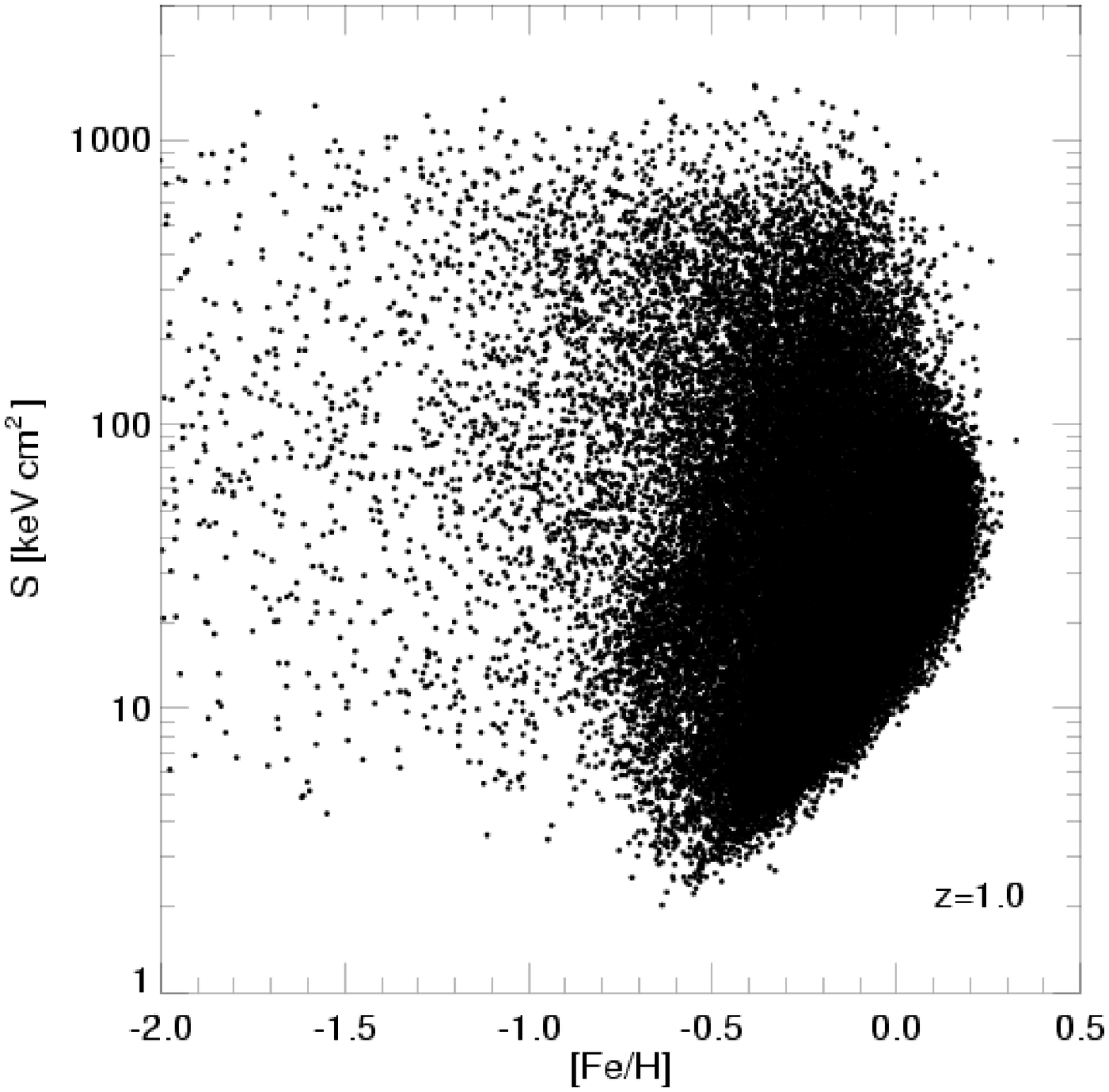}
  \end{minipage}%
  \begin{minipage}[c]{.33\textwidth}
    \centering \includegraphics[width=60mm]{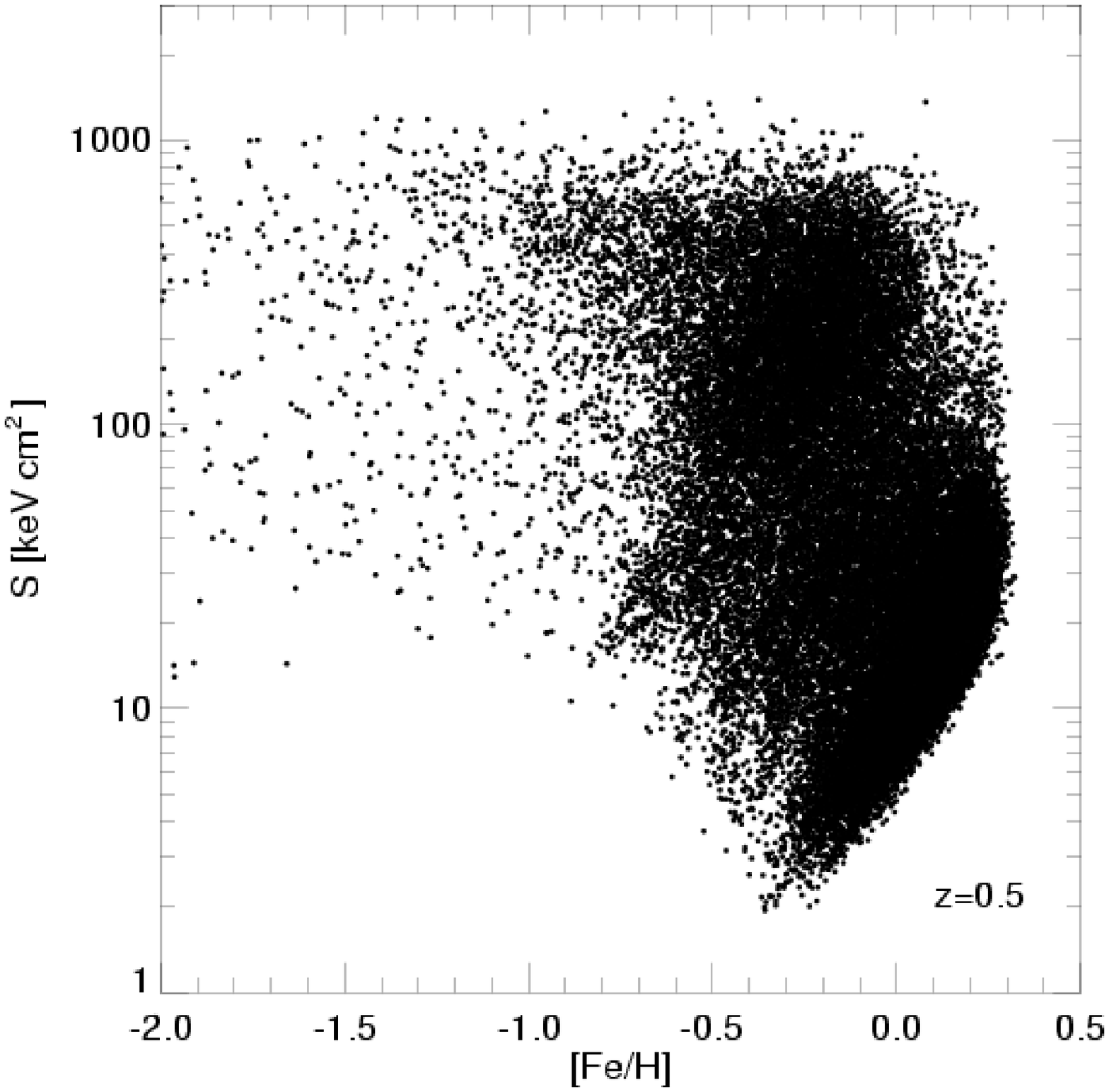}
  \end{minipage}%
  \begin{minipage}[c]{.33\textwidth}
    \centering \includegraphics[width=60mm]{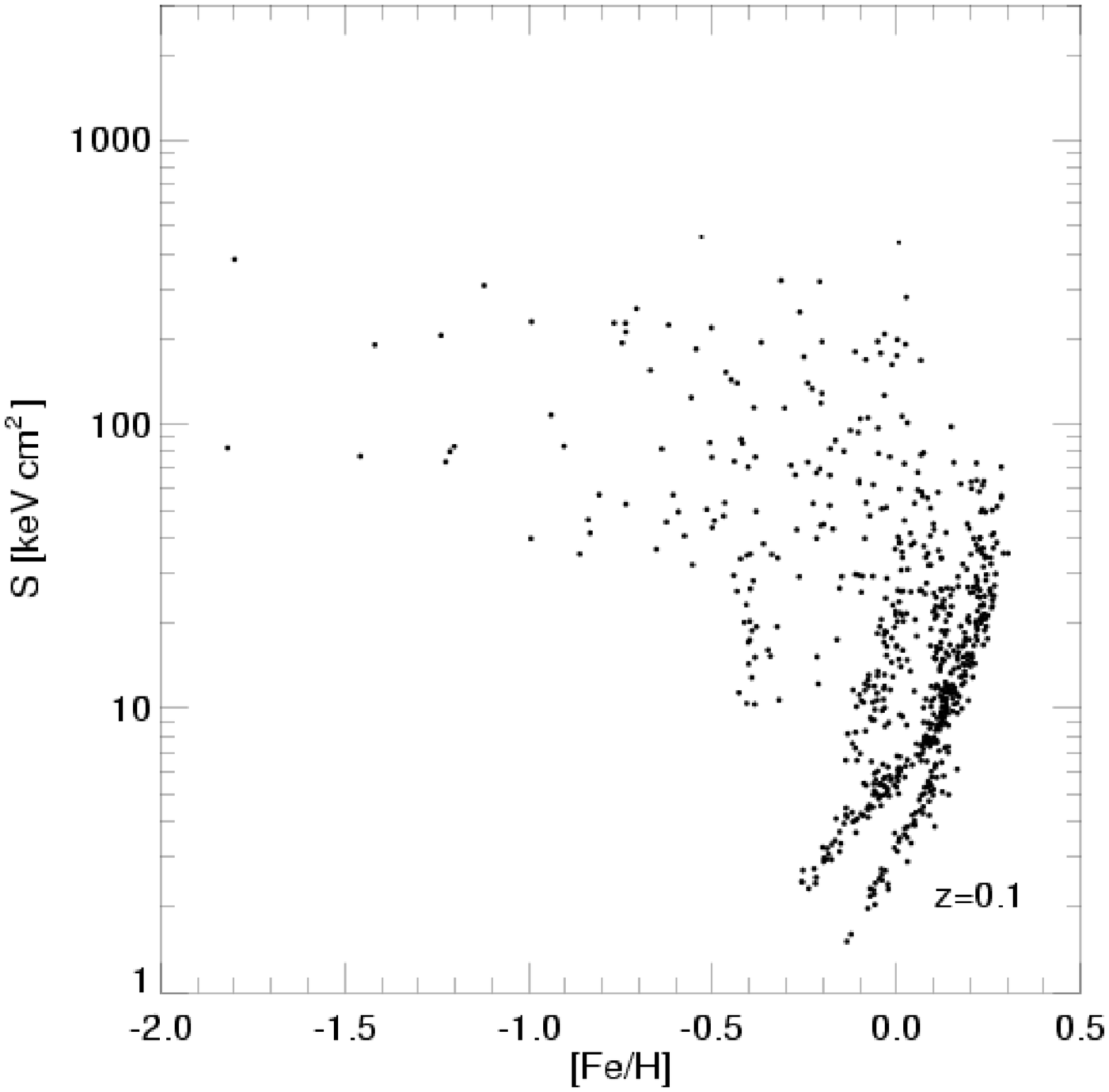}
  \end{minipage}%
\caption{
Evolution with redshift of the dependence between 
the entropy and iron abundance
of gas particles that 
are outside the virial radius of the main progenitor
of the cluster and end up within 
a sphere of radius
 $0.15 \,R_{\rm vir}$ centred at the cluster C1
at $z=0$.
Redshifts are indicated in each panel.
}
\label{fig6}
\end{figure*}

Figure \ref{fig4} shows the history of metal ejection since $z \la 1.4$.
For the simulated cluster C1,
we estimate the
accumulated 
iron mass ejected into the ICM below a given
redshift by galaxies contained within $0.05 \, R_{\rm vir}$ and $R_{\rm vir}$;
it is expressed in terms of the total mass of iron contained in the ICM
within those radii at $z=0$.
We see that, for $z \la 1.4$,
the 
accumulated 
iron mass ejected from galaxies within $R_{\rm vir}$
is $\sim 10$ per cent, while this fraction is reduced to  
$\sim 4 - 5$ per cent when only galaxies
within $0.15 \,R_{\rm vir}$ are taken into account.
In this last case, 
the ejected iron mass drops quite abruptly at $z \simeq 0.3$, 
making negligible contribution to the iron budget of the ICM within 
the innermost regions. 
We also show separately the contribution of different
types of supernovae. We find that $\sim 90$ per cent of the iron mass
ejected since $z < 1.4$ is provided by SNe Ia.

These values of mass of iron ejected can be considered as lower limits
because of the low star formation rate of cluster galaxies at low redshifts
in our model.
This behaviour arises as a result of the suppression of gas cooling in the
central galaxies
of halos with $V_{\rm vir} > 350 {\rm km}\,{\rm s}^{-1}$ and in satellite
galaxies; they simply form stars till their reservoir of cold gas is exhausted.
The first condition has been imposed to avoid too massive 
cluster central galaxies that are not consistent with observations.
The reduction or supression of the cooling flow could be naturally 
produced by AGN feedback
(\citealt{croton06}, \citealt{bower06}) which is not currently
included in our model.
As a result of the different modelling, 
it is not surprising to find that the iron ejected within 
$R_{\rm vir}$ since $z \la 0.5$
is a factor of two smaller than the estimations obtained
by the completely different approach of \citet{calura07}, based on
chemical evolution models
and morphological information, or those arising from self-consistent
cosmological hydrodynamical simulations \citep{tornatore07},
which has an excess of low-redshift star formation.

The fact that, at low redshift, the ICM in our model
receives a very small contribution of
iron ejected by cluster galaxies, 
allows us to isolate the
role
of dynamical processes in the evolution of the ICM abundance patterns.
With the aim of understanding the way
in which gas dynamics 
contributes to develop the iron abundance profiles,
we select gas particles located within a sphere of radius $0.15 \, R_{\rm vir}$
centred on the cluster at $z=0$.
They are tracked back to high redshifts in order to 
analyse the evolution of the 
relationship between their iron content and entropy values.
Figure~\ref{fig5} shows the dependence of such 
quantities
for cluster C1 at redshifts $z=0$, $0.5$ and $1$ for those
gas particles that remain within the main progenitor of the cluster 
as the redshift increases. 
Dots in the entropy-metallicity planes 
are colour-coded 
according to the distance of gas particles with respect to the 
centre of the main progenitor of the cluster: 
wihin $0.02\, R_{\rm vir}$ (red),
in the ranges $0.02 - 0.05 \, R_{\rm vir}$ (magenta), 
$0.05 - 0.15 \, R_{\rm vir}$ (blue),
$0.15 - 0.3 \, R_{\rm vir}$ (cyan) and $0.3 - 1 \, R_{\rm vir}$ 
(green).
The behaviuor of gas particles lying
outside the virial radius of the main progenitor
is presented in Figure~\ref{fig6},
at redshifts $z=0.1$, $0.5$ and $1$.

The distribution in the entropy-metallicity
plane of gas particles located at clustercentric distances
$\la 0.15\, R_{\rm vir}$ at $z=0$ (right panel of Figure~\ref{fig5})
simply 
reflecs the fact that the closer
are the particles from the cluster centre, the lower are their entropy values,
according to the entropy profile of the cluster \citep{dolag05}.
Only $\sim 19$ per cent of these particles resides at distances  
$\la 0.15\, R_{\rm vir}$ at $z\approx 1$
 (left panel of Figure \ref{fig5}),
while a larger percentage ($\sim 58$ per cent) is found outside the
main progenitor of the cluster (left panel of Figure \ref{fig6}).
This behaviour is consistent with the flattening of metallicity 
profiles as redshift increases (see Figure \ref{fig2}).
Both sets of gas particles, within and outside the virial radius of the
main progenitor, have already achieved a high level of enrichment 
by $z\approx 1$.
Around $14 - 15$ per cent of gas associated to other progenitor halos 
is characterized by low entropy values ($\la 10\,{\rm keV}\,{{\rm cm}^2}$)
which are not so common among gas particles lying within the main progenitor.

As the cluster assembles, gas particles outside the
main progenitor of the cluster
are incorporated into it following the infalling substructures, which
are stripped in the process.
At $z\approx 0.5$, the percentage of gas particles contained within the
virial radius of the
main progenitor increases to $\sim 63$ per cent, with $\sim 40$ lying at
$\la 0.15\, R_{\rm vir}$.
Very few particles are found outside
the main progenitor at $z\approx 0.1$, as can 
be seen from 
the right panel of Figure \ref{fig6}.
A small fraction of gas
is stripped quite early from the halo progenitors;
they are then incorporated onto the cluster by some diffusive accretion
mechanism,
increasing their entropy as a result of the shocks suffered during
the process \citep{dolag05}.
As time passes by, the enriched low entropy gas converges to the
inner regions of the cluster.
The entropy profile of the ICM
regulates this contribution.
Although we have non-radiative simulations, the temperature profile
of the ICM at $z=0$ decreases for radius smaller than
$0.05 \, R{\rm vir}$. 
As discussed by \citet{churazov03},
the presence of this dense and cool
core only allows the penetration
of the low entropy gas attached to the infalling dark matter
substructures that has been enriched at high redshift when
the iron ejecta from galaxies were larger.
The mass accreted onto the main progenitor
at a given time may vary among the different clusters analysed
because of their particular formation history; 
however, the general features of the evolution in the entropy-metallicity 
plane is the same for all
of them.

From this analysis, we find that much of the central, most highly enriched gas 
originates in infalling substructures that are not the main 
cluster progenitor.
This scenario of the origin of the iron abundance pattern
is supported by
the results of a test case 
in which gas particles are contaminated only by
galaxies contained within the main progenitor of the cluster.
In such a restricted enrichment scheme, 
the iron abundances are very low and the shape of the
profile becomes quite flat in comparison with the one obtained when the
contribution of all galaxies is taken into account.
The dynamical
process that involves the sinking of enriched low entropy gas explains the
evolution suffered by the metallicity profiles in their central parts,
which cannot be accounted for by recent enhanced star formation activity.
We can see the steepening of the profiles as a consequence of the 
spatial redistribution of the intracluster gas with different level of 
chemical enrichment.

\section{Conclusions}\label{sec_conclu}

We have applied an hybrid model for metal enrichment of the ICM
that combines hydrodynamical cosmological simulations of galaxy clusters
and a semi-analytic model of galaxy formation. This hybrid model has the
special feature of linking the metal production of galaxies in the
semi-analytic model with the chemical enrichment of gas particles
in the {\em N}-body/SPH
simulations. This allows us to follow the evolution with time
of the spatial distribution of metals in the intracluster gas.

We have adopted a
scheme for metal spreading among gas particles which
is based on the dynamics of the expansion of supernova-driven
superbubbles generated around galaxies \citep{bertone05}. 
We find that this physically motivated estimation of the 
enrichment radius 
satisfies observed ICM metallicity 
properties both locally and at high redshifts,
being in good agreement with iron abundance profiles at $z=0$
and the evolution of the mean
abundances with lookback time.
The simulated mean metallicities within
$0.15 \, R{\rm vir}$ are in good agreement 
with the observations analysed by \citet{balestra07} in the whole
redshift range considered, closely following the 
parametrization of 
observed data till $z \la 1.4$.
The results of this model let us conclude that:

\begin{enumerate}
\item 
The observed mean metallicity evolution \citep{balestra07}
can be explained as being
produced by the combination of
an overall increase of the iron content
within the virial radius of the main progenitor of the cluster, and 
the evolution
of the slope of the
central part of the iron abundance profiles ($\la 0.15  \, R_{\rm vir}$).
This result has been obtained from the analysis of both MW and
EM iron abundance profiles 
and mean metallicities,
with EW mean values mainly providing information of the central parts 
of the clusters.
\item
The metallicities of gas particles that at $z=0$ end up 
within a region  delimited by a radius of $\sim 0.15  \, R_{\rm vir}$
have been already achieved between $z \sim 1$ and $\sim 0.5$, 
when almost half of the gas has not yet been 
accreted onto the main progenitor of the cluster, and
the metal ejecta from galaxies was considerably higher with respect to
the present epoch.
\item
As time passes by, the enriched low entropy gas,
mainly attached to infalling substructures, sinks into
the cluster and is mixed there contributing
to develop the iron abundance profiles.
The presence of a dense and cool
core allows the penetration
of this low entropy gas 
\citep{churazov03}.
Hence, the very central metallicity ($\la 0.02  \, R_{\rm vir}$)  
originates almost exclusively from gas in infalling substructures 
that converge to
the cluster core.
This dynamical process explains the
evolution suffered by the metallicity profiles in their central parts,
which cannot be accounted for by recent enhanced star formation activity.
\end{enumerate}

The shape of entropy profiles determine the development of the
abundance profiles. The forecoming conclusions have been obtained
for non-radiative simulations of galaxy clusters. The entropy profile
change substantially for radiative runs, as has been 
shown by \citet{borgani05}, with the main change occurring
within $\sim 0.1 \, R{\rm vir}$. 
For simulations that include gas cooling, star formation
and feedback regulated by the velocity of galactic winds,
the entropy profile monotonically decreases to the cluster
centre, while in non-radiative simulations an almost isentropic core develops.
The change in the central entropy values would only affect the limit on the
entropy of the gas in infalling substructures that is allowed to penetrate 
the core,
without affecting the general development of the profiles that we have 
described by the analysis of our results.

Gas cooling, star formation and feedback
processes from supernovae explosions associated to galaxy formation
are taken into account in the
semi-analytic model.
The lack of AGN feedback in our model oblige us to use simple rules
for the quenching of star formation to avoid overcooling in the semi-analytical
prescription, and therefore we do not have a robust control of the cooled
gas fraction.
The impact of AGN feedback on galaxy properties is currently being
investigated (Lagos, Cora \& Padilla, in prep.); this improved
semi-anlytic model will be used to
study the influence of AGN feedback on the ICM metal enrichment.

In the case of cool-core clusters,
the formation
and evolution of the central brightest galaxy plays an important
role
(\citealt{bohringer04, degrandi04}).
Its effect will be properly modelled  with the inclusion
of AGN feedback that mainly affects star formation in masive halos;
this will allow to evaluate its influence on the development of the ICM
metallicity, and to quantify the relative importance with respect
to gas dynamical effects.

The semi-analytic model does not consider tidal disruption of cluster
galaxies which give raise to intracluster stars, as shown by
high resolution {\em N}-body/SPH simulations
(\citealt{murante04, sommerlarsen05}). 
In our model,
the potential intracluster stars remain within the galaxies; their
ejected supernovae products are affected by the feedback process considered
instead of being injected
directly into the intracluster gas, the latter being the main
advantange of the unbound component in the enrichment process of the ICM \citep{zaritsky04}.
Thus, we can consider that our model 
underestimates the contribution of potential
intracluster stars to the metal
budget of the ICM. 
On the other hand, 
numerical simulations show that unbound stars (at least $\sim 10$~per cent
of the cluster stellar mass)
accumulate within the cluster generating intracluster light
with a shallower radial profile than that of the bound component
within $\sim 0.3\,R_{\rm vir}$; besides, they 
are on average older than the stars in cluster galaxies,
which iron ejection rates already peak at high redshift.
Hence, based on these results, we do not expect that intracluster 
stars drive a significant
evolution of the ICM iron abundance profile at low redshifts.

Ram pressure stripping is another process that might change the 
shape of the slope of
the abundance pattern in the innermost part of the ICM.
Its effect is present on the
least bound gas particles in the hydrodynamical simulations.
However, our semi-analytic model does not currently
include the ram pressure stripping
on the cold gas of cluster galaxies, which would affect the cold gas
reservoir available for star formation and the timing of galactic mass loss.
The evaluation of the impact of this mechanism will be part of future
work.

\section*{Acknowledgments}
We warmly thank Stefano Borgani, who proposed this research project, for
largely contributing to this work with useful suggestions and comments.
We acknowledge the annonymous referee for helpful remarks
that allow to 
improve this work.
S.A.C. is very grateful for the
hospitality of the Osservatorio Astronomico di Trieste, where this
project was started, and the travel grant from the National Institute
for Nuclear Physics, Trieste, Italy. 
P.T. acknowledges
financial contribution from contract ASI--INAF I/023/05/0 and from the
PD51 INFN grant.
We thank Fabio Gastaldello
for helpful comments. 
The simulations were carried out with CPU time allocated at the
``Centro Interuniversitario del Nord-Est per il Calcolo Elettronico"
(CINECA, Bologna), thanks to grants of INAF and from the University of Trieste.
This project was
financially supported by Fundaci\'on Antorchas, PIP 5000/2005 from
Consejo Nacional de Investigaciones Cient\'{\i}ficas y T\'ecnicas,
and PICT 26049 of Agencia de Promoci\'on Cient\'{\i}fica y T\'ecnica, Argentine.

\bsp

\label{lastpage}


\begin{thebibliography}{99}

\bibitem[\protect\citeauthoryear{Anders \& Grevesse}{1989}]{anders89}
Anders E., Grevesse N., Geochim. Cosmochim. Acta, 53, 197
Proceedings 183,
p. 1, Ed. C.J. Waddington (New York: AIP)
\bibitem[\protect\citeauthoryear{Asplund et al.}{2005}]{asplund05}
Asplund M., Grevesse N., Sauval A. J., 2005, 
in Cosmic Abundances as Records of Stellar Evolution and Nucleosynthesis, 
ed. T.G. Barnes III, F.N. Bash, ASP Conference Series, 336, 25 
\bibitem[\protect\citeauthoryear{Balestra et al.}{2007}]{balestra07}
Balestra I., Tozzi P., Ettori S., Rosati P., Borgani S., Mainieri V., Norman C., Viola M., 2007, A\&A, 462, 429 
\bibitem[\protect\citeauthoryear{Bertone et al.}{2005}]{bertone05}
Bertone S., Stoehr F., White S.D.M., 2005, MNRAS, 359, 1201
\bibitem[\protect\citeauthoryear{B\"ohringer et al.}{2004}]{bohringer04}
B\"ohringer H., Matsushita K., Churazov E., Finoguenov A., Ikebe Y., 2004, A\&A, 416, L21
\bibitem[\protect\citeauthoryear{Borgani et al.}{2005}]{borgani05}
Borgani S., Finoguenov A., Kay S.T., Ponman T.J., Springel V.,
Tozzi P., Voit G.M., 2005, MNRAS, 361, 233
\bibitem[\protect\citeauthoryear{Bower et al.}{2006}]{bower06}
Bower R.G., Benson A.J., Malbon R., Helly J.C., Frenk C.S., Baugh C.M.,
Cole S., Lacey C.G., 2006, MNRAS, 370, 645
\bibitem[\protect\citeauthoryear{Calura et al.}{2007}]{calura07}
Calura F., Matteucci F., Tozzi P., 2007, MNRAS, 378L, 11 
\bibitem[\protect\citeauthoryear{Cora}{2006}]{cora06}
Cora S.A., 2006, MNRAS, 368, 1540
\bibitem[\protect\citeauthoryear{De Lucia et al.}{2004}]{lucia04}
De Lucia G., Kauffmann G., White S.D.M., 2004, MNRAS, 349, 1101
\bibitem[\protect\citeauthoryear{Dolag et al.}{2005}]{dolag05}
Dolag K., Vazza F., Brunetti G., Tormen G.G., 2005, MNRAS, 364, 753
\bibitem[\protect\citeauthoryear{Croton et al.}{2006}]{croton06}
Croton D.J., Springel V., White S.D.M., De Lucia G., Frenk C.S.,
Gao L., Jenkins A., Kauffmann G., Navarro J.F., Yoshida N., 
2006, MNRAS, 365, 11	
\bibitem[\protect\citeauthoryear{Churazov et al.}{2003}]{churazov03}
Churazov E., Forman W., Jones C., B\"ohringer H., 2003, ApJ, 590, 225
\bibitem[\protect\citeauthoryear{De Grandi \& Molendi}{2001}]{degrandi01}
De Grandi S., Molendi S., 2001, ApJ, 551, 153
\bibitem[\protect\citeauthoryear{De Grandi et al.}{2004}]{degrandi04}
De Grandi S., Ettori S., Longhetti M., Molendi S. 2004, A\&A, 419, 7
\bibitem[\protect\citeauthoryear{Erb et al.}{2006}]{erb06}
Erb D.K., Shapley A.E., Pettini M., Steidel C.C., Reddy N.A., Adelberger K.L., 2006, ApJ, 644, 813
\bibitem[\protect\citeauthoryear{Ettori}{2005}]{ettori05}
Ettori S., 2005, MNRAS, 262, 110
\bibitem[\protect\citeauthoryear{Greggio \& Renzini}{1983}]{greggio83} 
Greggio L., Renzini A., 1983, A\&A, 118, 217
\bibitem[\protect\citeauthoryear{Heckman et al.}{2000}]{heckman00}
Heckman T.M., Lehnert M.D., Strickland D.K., Armus L.,
2000, ApJS, 129, 493
\bibitem[\protect\citeauthoryear{Iwamoto et al.}{1999}]{Iwamoto99}
Iwamoto K., Brachwitz F., Nomoto K., Kishimoto N., Umeda H., Hix W.H.,
Thielemann F.-K., 1999, ApJS, 125, 439
\bibitem[\protect\citeauthoryear{Kapferer et al.}{2007}]{kapferer07}
Kapferer W., Kronberger T., Weratschnig J., Schindler S., Domainko W.,
van Kampen E., Kimeswenger S., Mair M., Ruffert M., 2007, MNRAS,
466, 813 
\bibitem[\protect\citeauthoryear{Lia, Portinari \& Carraro}{2002}]{lia02}
Lia C., Portinari L., Carraro G., 2002, MNRAS, 330, 821
\bibitem[\protect\citeauthoryear{Loewenstein}{2001}]{loewenstein01}
Loewenstein M., 2001, ApJ, 557, L573
\bibitem[\protect\citeauthoryear{Loewenstein}{2006}]{loewenstein06}
Loewenstein M., 2006, ApJ, 648, 230 
\bibitem[\protect\citeauthoryear{Marigo}{2001}]{marigo01}
Marigo P., 2001, A\&A, 194, 217
\bibitem[\protect\citeauthoryear{Maughan et al.}{2008}]{maughan08}
Maughan B.J., Jones C., Forman W., Van Speybroeck L., 2008, ApJS, 174, 117
\bibitem[\protect\citeauthoryear{Moll et al.}{2006}]{moll07}
Moll R., Schindler S., Domainko W., Kapferer W., Mair M., van Kampen E., Kronberger T., Kimeswenger S., Ruffert M. 2007, A\&A, 463, 513
\bibitem[\protect\citeauthoryear{Murante et al.}{2004}]{murante04}
Murante G., Arnaboldi M., Gerhard O., Borgani S., Cheng L.M., Diaferio A., Dolag K., Moscardini L., Tormen G., Tornatore L., Tozzi P., 2004, ApJ, 607, L83
\bibitem[\protect\citeauthoryear{Portinari et al.}{1998}]{portinari98}
Portinari L., Chiosi C., Bressan A., 1998, A\&A, 334, 505
\bibitem[\protect\citeauthoryear{Nagashima et al.}{2005}]{nagashima05}
Nagashima M., Lacey C., Baugh C.M., Frenk C.C., 2005, MNRAS, 358, 1247
\bibitem[\protect\citeauthoryear{Padovani \& Matteucci}{1993}]{padovani93}
Padovani P., Matteucci F., 1993, ApJ, 416,26
\bibitem[\protect\citeauthoryear{Rebusco et al.}{2005}]{rebusco05}
Rebusco P., Churazov E., B\"ohringer H., Forman W., 2005, MNRAS, 359, 1041
\bibitem[\protect\citeauthoryear{Recchi et al.}{2001}]{recchi01}
Recchi S., Matteucci F., D'Ercole A., MNRAS, 322, 800 
\bibitem[\protect\citeauthoryear{Renzini}{1997}]{renzini97}
Renzini A., 1997, ApJ, 488, 35
\bibitem[\protect\citeauthoryear{Roediger et al.}{2007}]{roediger07}
Roediger E., Br\"uggen M., Rebusco P., B\"ohringer H., Churazov E., 2007, MNRAS, 375, 15 
\bibitem[\protect\citeauthoryear{Romeo et al.}{2006}]{romeo06}
Romeo A.D., Sommer-Larsen J., Portinari L., Antonuccio-Delogu V., 2006, MNRAS, 371, 548
\bibitem[\protect\citeauthoryear{Sommer-Larsen, Romeo \& Portinari}{2005}]{sommerlarsen05}
Sommer-Larsen J. Romeo A.D., Portinari L., 2005, MNRAS, 357, 478
\bibitem[\protect\citeauthoryear{Springel, Yoshida \& White}{2001}]{springel01a}
Springel V., Yoshida N., White S.D.M., 2001, New Astronomy, 79, 117
\bibitem[\protect\citeauthoryear{Springel et al.}{2001}]{springel01}
Springel V., White S.D.M., Tormen G., Kauffmann G., 2001, MNRAS, 238, 726
\bibitem[\protect\citeauthoryear{Springel \& Hernquist}{2002}]{springel02}
Springel V., Hernquist L., 2002, MNRAS, 333, 649 
\bibitem[\protect\citeauthoryear{Springel \& Hernquist}{2003}]{springel03}
Springel V., Hernquist L., 2003, MNRAS, 339, 289 
\bibitem[\protect\citeauthoryear{Springel}{2005}]{springel05}
Springel V., 2005, MNRAS, 364, 1105 
\bibitem[\protect\citeauthoryear{Sutherland \& Dopita}{1993}]{suth93}
Sutherland R.S., Dopita M.A., 1993, ApJS, 88, 253
\bibitem[\protect\citeauthoryear{Tamura et al.}{2004}]{tamura04}
Tamura T., Kaastra J.S., Herder J.W.A.den., Bleeker J.A.M., Peterson J.R., 2004,
submitted to A\&A, 420, 135 
\bibitem[\protect\citeauthoryear{Tormen et al.}{1997}]{tormen97}
Tormen G. Bouchet F, White S.D.M., 1997, MNRAS, 286, 865
\bibitem[\protect\citeauthoryear{Tornatore et al.}{2007}]{tornatore07}
Tornatore L., Borgani S., Dolag K., Matteucci F.,
2007, MNRAS, 382, 1050 
\bibitem[\protect\citeauthoryear{Valdarnini}{2003}]{valdarnini03} 
Valdarnini R. 2003, MNRAS, 339, 1117
\bibitem[\protect\citeauthoryear{Vikhlinin et al.}{2006}]{vikhlinin06}
Vikhlinin A., Kravtsov A., Forman W. Jones C., Markevitch M., Murray S.S., Van Speybroeck L., 2006, ApJ, 640, 691
\bibitem[\protect\citeauthoryear{Vikhlinin et al.}{2005}]{vikhlinin05}
Vikhlinin A., Markevitch M., Murray S.S., Jones C., Forman W., Van Speybroeck L., 2005, ApJ, 628, 655
\bibitem[\protect\citeauthoryear{Yoshida et al.}{2001}]{yoshida01}
Yoshida N., Sheth R.K., Diaferio A., 2001, MNRAS, 328, 669
\bibitem[\protect\citeauthoryear{Zaritsky, Gonzalez \& Zabludoff}{2004}]{zaritsky04}
Zaritsky D., Gonzalez A.H., Zabludoff A.I., 2004, ApJ, 613, L93

\end{thebibliography}
\end{document}